\documentclass[a4paper,12pt]{article}

\begin{document}

\def\chic#1{{\scriptscriptstyle #1}}
\newcommand{\be}{\begin{equation}}
\newcommand{\bea}{\begin{eqnarray}}
\newcommand{\ee}{\end{equation}}
\newcommand{\eea}{\end{eqnarray}}

\setcounter{equation}{0}
\def\theequation{\arabic{section}.\arabic{equation}}

\begin{flushright}
FTUV-01-02-13\\
IFIC/01-6
\end{flushright}
 
\begin{center}
{\Large {\bf Breit-Wigner formalism\\[.4cm]
for non-Abelian theories}}
\\[2.4cm]
{\large Joannis Papavassiliou} 
\\[0.4cm]
{\em 
Departamento de F\'{\i}sica Te\'orica and IFIC, 
Universidad de Valencia \\
E-46100 Burjassot (Valencia), Spain}\\[0.3cm]
\end{center}

\begin{abstract}
 {\normalsize   The  consistent  description of
resonant  transition amplitudes within  the framework  of perturbative
field  theories   necessitates  the  definition   and  resummation  of
off-shell  Green's  functions,  which  must  respect  several  crucial
physical requirements.  In particular, the generalization of the usual
Breit-Wigner formalism  in a non-Abelian context  constitutes a highly
non-trivial  problem,  related to  the  fact  that the  conventionally
defined Green's functions are  unphysical.  We briefly review the main
field-theoretical  difficulties  arising when  attempting  to use  such
Green's functions  outside the confines of a  fixed order perturbative
calculation,  and   explain  how   this  task  has   been  successfully
accomplished in the framework of the pinch technique.
\vspace*{0.4 cm} }
\end{abstract}

\vspace{3.0cm}
\begin{center}
{Contribution to A.Sirlin's seventieth birthday
Symposium \\ 
``50 Years of Electroweak
Precision Physics'',\\ 
New York University,
October 27-28, 2000} .
\end{center}

\newpage

\setcounter{equation}{0}
\section{ The Breit-Wigner resummation} 

The  physics of  unstable particles  and the  computation  of resonant
 transition amplitudes  has attracted significant  attention in recent
 years,   because   it  is   both   phenomenologically  relevant   and
 theoretically  challenging.   Throughout  the nineties  A.Sirlin  and
 collaborators  \cite{AS}  have  addressed  various  important  issues
 related to  the proper  definition of masses  and widths  of unstable
 particles. During  the same period  A.Pilaftsis and I  have developed
 the  formalism which  allows  for the  proper  generalization of  the
 Breit-Wigner resummation in a non-Abelian context \cite{PP2,PPHiggs}.
 In  what  follows  I   will  outline  the  conceptual  and  practical
 difficulties appearing when  dealing with non-Abelian resonances, and
 will briefly  explain how the  field-theoretical method known  as the
 pinch   technique  (PT)   \cite{PT,PT2}  allows   for   a  consistent
 description of resonant transition amplitudes.

The mathematical  expressions for computing  transition amplitudes are
ill-defined  in the  vicinity  of resonances,  because the  tree-level
propagator of  the particle  mediating the interaction,  i.e. $\Delta=
(s-M^2)^{-1}$,   becomes  singular   as   the  center-of-mass   energy
$\sqrt{s}\sim  M$.   The standard  way  for  regulating this  physical
kinematic  singularity is to  use a  Breit-Wigner type  of propagator,
which  essentially  amounts   to  the  replacement  $(s-M^2)^{-1}  \to
(s-M^2+iM\Gamma)^{-1}$, where  $\Gamma$ is  the width of  the unstable
(resonating)  particle. The  field-theoretic  mechanism which  enables
this  replacement   is  the   Dyson  resummation  of   the  (one-loop)
self-energy  $\Pi(s)$ of  the unstable  particle, which  leads  to the
substitution $(s-M^2)^{-1} \to [s-M^2+\Pi(s)]^{-1}$; the running width
of the particle is then defined as $M\Gamma(s) =\Im m \Pi(s)$, whereas
the usual (on-shell) width is simply its value at $s=M^2$.

It is  well-known that,  to any {\it  finite order},  the conventional
perturbative  expansion   gives  rise  to   expressions  for  physical
amplitudes which are endowed with crucial properties. For example, the
amplitudes are independent of  the gauge-fixing parameter (GFP) chosen
to  quantize the  theory, they  are gauge-invariant  (in the  sense of
current conservation),  they are unitary (in the  sense of probability
conservation), and well behaved at high energies. The above properties
are however  not always encoded into the  individual Green's functions
which  are  the building  blocks  of  the aforementioned  perturbative
expansion; indeed,  the simple fact  that Green's functions  depend in
general explicitly  on the  GFP, indicates that  they are void  of any
physical  meaning.  Evidently,  when  going  from  unphysical  Green's
functions to  physical amplitudes subtle  field-theoretical mechanisms
are at  work, which  implement highly non-trivial  cancellations among
the various Green's functions appearing at a given order.

The happy  state of affairs  described above is guaranteed  within the
framework  of the conventional  perturbative expansion,  provided that
one works  at a given fixed  order.  It is relatively  easy to realize
however that  the Breit-Wigner  procedure is in  fact equivalent  to a
reorganization  of  the  perturbative  series; indeed,  resumming  the
self-energy  $\Pi$ amounts to  removing a  particular piece  from each
order of the perturbative expansion, since from all the Feynman graphs
contributing to a given order $n$  we only pick the part that contains
$n$ self-energy bubbles $\Pi$, and then take $n \to \infty$.  However,
given that a non-trivial cancellation involving the unphysical Green's
function  is  generally  taking  place  at  any  given  order  of  the
conventional perturbative  expansion, the act of removing  one of them
from each order may or may  not distort those cancellations. To put it
differently, if  $\Pi$ contains unphysical  contributions (which would
eventually cancel against other pieces within a given order) resumming
it naively  would mean that  these unphysical contributions  have also
undergone infinite  summation (they now  appear in the  denominator of
the propagator $\Delta$).  In order to remove them one  has to add the
remaining  perturbative  pieces  to  an  infinite  order,  clearly  an
impossible  task,  since  the  latter  (boxes  and  vertices)  do  not
constitute a resumable set.  Thus,  if the resumed $\Pi$ happened to
contain such unphysical terms, one would finally arrive at predictions
for  the physical  amplitude close  to  the resonance  which would  be
plagued with unphysical artifacts.  It turns out that, while in scalar
field theories and Abelian gauge  theories $\Pi$ does not contain such
unphysical  contributions, this  seizes  to  be true  in  the case  of
non-Abelian gauge theories.

The  most obvious  signal  revealing that  the conventionally  defined
non-Abelian  self-energies  are not  good  candidates for  resummation
comes from  the simple calculational  fact that the  bosonic radiative
corrections  to the self-energies  of vector  ($\gamma$, $W$,  $Z$) or
scalar (Higgs) bosons induce a  non-trivial dependence on the GFP used
to define the tree-level  bosonic propagators appearing in the quantum
loops.  This is  to be contrasted to the  radiative corrections due to
fermion  loops,  which,  even  in  the context  of  non-Abelian  gauge
theories behave as in  quantum electrodynamics (QED), {\em i.e.}, they
are    GFP-independent.    In    addition,    formal   field-theoretic
considerations as  well as direct calculations show  that, contrary to
the QED case,  the non-Abelian Green's functions do  not satisfy their
naive, tree-level Ward identities , after bosonic one-loop corrections
are  included.   A  careful   analysis  shows  that  this  fundamental
difference between  Abelian and non-Abelian  theories has far-reaching
consequences; the  naive generalization of the  Breit-Wigner method to
the latter case  gives rise to Born-improved amplitudes,  which do not
faithfully  capture the  underlying  dynamics.  Most  notably, due  to
violation of the optical theorem, unphysical thresholds and artificial
resonances  appear, which  distort the  line-shapes of  the resonating
particles.  In addition, the high energy properties of such amplitudes
are  altered,  and are  in  direct  contradiction  to the  equivalence
theorem (ET) \cite{EqTh}.

In  order  to  address  these  issues,  a  new  approach  to  resonant
 transition  amplitudes has  been developed  over the  past  few years
 \cite{PP2,PPHiggs},  which  is  based  on  the  the  pinch  technique
 (PT)~\cite{PT,PT2}; the  latter is  a diagrammatic method  whose main
 thrust is to exploit the symmetries built into physical amplitudes in
 order to  construct off-shell sub-amplitudes  which are kinematically
 akin to  conventional Green's functions, but, unlike  the latter, are
 also endowed with several crucial properties:
\begin{itemize}
\item[(a)] They are independent of  the GFP, within any
gauge-fixing scheme chosen to quantize the theory.

\item[(b)] They satisfy naive (ghost-free) tree-level  Ward identities  
instead of the usual Slavnov-Taylor  identities.     

\item[(c)] They display    physical
thresholds  only \cite{PP2}.

\item[(d)]
They  satisfy  individually the optical  and
equivalence theorems \cite{PP2,PPHiggs}. 

\item[(e)] 
The   effective two-point functions  constructed  are
universal (process-independent) ,  
Dyson-resumable \cite{PP2},  and do not  shift
the  position of the  gauge-independent complex  pole \cite{PP2}.
In addition, one may use them to 
construct ``effective charges'', i.e process-independent
and renormalization-group-invariant objects \cite{PPHiggs}.

\item[(f)]  The PT effective Green's functions {\it coincide}
with the conventional Green's functions defined in the
framework of the background field method \cite{BFM}, 
when the latter are
computed in the Feynman gauge \cite{DDW}.

\end{itemize}

The crucial novelty introduced by the PT is that
the resummation of graphs must take place only
{\em after} the amplitude of interest has been cast
via the PT algorithm into
manifestly physical sub-amplitudes, with distinct
kinematic properties, order by order in perturbation theory.
Put in the language employed earlier, the PT 
ensures that all unphysical contributions contained inside $\Pi$ have
been identified and properly discarded, before $\Pi$ undergoes 
resummation. 
It is important to emphasize that the only ingredient needed for
constructing the PT effective Green's functions is the full exploitation
of elementary Ward-identities , which are a direct consequence
of the BRS \cite{BRS} symmetry of the theory,
and the proper use of the unitarity
and analyticity of the $S$-matrix.
In what follows I will describe some of the salient features 
of this method.

\section{ The Pinch Technique.}

 Within the PT framework, the transition
amplitude $T(s,t,m_i)$ of a $2\to 2$ process,
can be decomposed as
\begin{equation}
\label{TPT}
T(s,t,m_i)\ =\ \widehat{T}_1(s)\ +\ \widehat{T}_2(s,m_i)\ +\
\widehat{T}_3(s,t,m_i),
\end{equation}
in terms of three individually gauge-invariant quantities:
a propagator-like part ($\widehat{T}_1$), a vertex-like piece
($\widehat{T}_2$),
and a part containing box graphs ($\widehat{T}_3$). The important observation
is that vertex and box graphs contain in general
pieces, which are kinematically akin to self-energy graphs
of the transition amplitude (Fig.1)
The PT is a systematic way of extracting such pieces and
appending them to the conventional self-energy graphs.
In the same way, effective gauge invariant
vertices may be constructed, if
after subtracting from the conventional vertices the
propagator-like pinch parts we add the vertex-like pieces coming from
boxes. The remaining purely box-like contributions are then
also gauge invariant. The way to identify the pieces which are
to be reassigned, all one has to do is to resort to the fundamental
PT cancellation, which is in turn a direct consequence of the
elementary Ward identities of the theory. 

There are two basic ingredients in the PT construction.
The first is the identification of all longitudinal momenta
involved, i.e. the momenta which can trigger the elementary Ward identities.
There are two sources of such momenta: The tree-level expressions
for the gauge boson propagators appearing inside Feynman diagrams
and the tri-linear gauge boson vertices.
For example, in the case of QCD, the tree-level gluon propagator reads
\be
i\Delta_{\mu\nu}(q) = \frac{-i}{q^2}\bigg(g_{\mu\nu} -
(1-\xi)\frac{q_{\mu}q_{\nu}}{q^2}\bigg)\, , 
\ee
and the longitudinal momenta are simply those multiplying 
$(1-\xi)$. The identification of the longitudinal momenta stemming
from the three-gluon vertex is slightly more subtle. To see how they emerge,
one must split the Bose-symmetric three-gluon vertex in the
following Bose-asymmetric way:    
\begin{eqnarray}
\Gamma_{\lambda\mu\nu} (q,-k_1,-k_2) &=& 
(k_1-k_2)_{\lambda}g_{\mu\nu}\, +\, (q+k_2)_{\mu}g_{\lambda\nu}
-\, (q+k_1)_{\nu}g_{\lambda\nu}\nonumber\\
 &=& [(k_1-k_2)_{\lambda}g_{\mu\nu}+
2q_{\mu}g_{\lambda\nu}-2q_{\nu}g_{\lambda\mu}]
+\ (k_{2\nu}g_{\lambda\mu}-k_{1\mu}g_{\lambda\nu}) \nonumber\\
&=& \Gamma^F_{\lambda\mu\nu}(q,-k_1,-k_2)\, + 
\Gamma^P_{\lambda\mu\nu}(q,-k_1,-k_2)\, .
\label{GFGP}
\end{eqnarray}
The $\Gamma^P_{\lambda\mu\nu}(q,-k_1,-k_2)$ contains the
aforementioned longitudinal momenta. 
The vertex  $\Gamma^F_{\lambda\mu\nu}(q,-k_1,-k_2)$
is Bose-symmetric only with respect to the
$\mu$ and $\nu$ legs.
The
first term in  $\Gamma^F_{\lambda\mu\nu}(q,-k_1,-k_2)$ is a convective
vertex describing the coupling of a gluon to a scalar field, 
whereas the
other two terms originate from gluon spin or magnetic moment. 
Evidently the above decomposition assigns a special r\^ole 
to the $q$-leg,
and allows $\Gamma^F_{\lambda\mu\nu}(q,-k_1,-k_2)$
to satisfy the elementary Ward identity
\begin{equation}
q^{\mu}\Gamma^{F}_{\mu\alpha\beta}(q, -k_1, -k_2)\ =\ 
(k_1^2\ -\ k_2^2)g_{\alpha\beta}\, .
\label{FWI}
\end{equation}

The second PT ingredient is the following: One has
to use all longitudinal momenta
identified above in order to trigger a fundamental, BRS-driven
cancellation involving graphs of different kinematic dependence. 
In particular, let us consider the amplitude
$q {\bar q}\rightarrow g g $, to be denoted by
${\cal T}=\langle q\bar{q}|T|gg\rangle$.
Diagrammatically, the amplitude ${\cal
T}$ consists of two distinct parts: $t$ and $u$-channel graphs that contain an
internal quark propagator, ${{\cal T}_{t}}^{ab}_{\mu\nu}$, as shown in Figs.\
4(d), and an $s$-channel amplitude, ${{\cal T}_{s}}^{ab}_{\mu\nu}$,
which is given in Fig.\ 4(a). The subscript ``$s$'' and ``$t$'' refers to the
corresponding Mandelstam variables, {\em i.e.}\ $s=q^2=
(p_1+p_2)^2=(k_1+k_2)^2$, and $t=(p_1-k_1)^2=(p_2-k_2)^2$. 
Specifically,
\begin{equation}
{\cal T}^{ab}_{\mu\nu}={{\cal T}_{s}}^{ab}_{\mu\nu}+
{{\cal T}_{t}}^{ab}_{\mu\nu}\, ,
\label{DefT}
\end{equation}
with
\begin{eqnarray}
{{\cal T}_{s}}^{ab}_{\mu\nu} & =&
-g^2 \bar{v}(p_2)\, \frac{\lambda^c}{2}\gamma_{\rho}\, u(p_1)
f^{abc}\, \bigg(\frac{1}{q^2}\bigg)
\Gamma_{\rho\mu\nu}(q,-k_1,-k_2)\, ,
\nonumber\\
{{\cal T}_{t}}^{ab}_{\mu\nu} &=& -g^2\bar{v}(p_2)\Big( 
\, \frac{\lambda^b}{2}\gamma^{\nu}\, S(\not\! p_1-\not\! k_1 )
\, \frac{\lambda^a}{2}\gamma^{\mu}\ +
\frac{\lambda^a}{2}\gamma^{\mu}\,
S(\not\! p_1-\not\! k_2)\, 
\gamma^{\nu}\frac{\lambda^b}{2}\, \Big)u(p_1) 
\nonumber\\
&&{}
\label{Tt}
\end{eqnarray}
where
\be
S(p) =\frac{i}{\not\! p - m}
\ee
is the quark propagator, and  $\lambda$ are the Gell-Mann matrices. 
It is elementary to verify that the action of the
longitudinal momenta $k_1^{\mu}$ or $k_2^{\nu}$ leads to
a non-trivial cancellation between the ${{\cal T}_{s}}$ and
${{\cal T}_{t}}$ amplitudes, as shown in Fig.4, which
is a direct consequence of the BRS symmetry.
Its one-loop implementation necessitates only the use of the
following basic Ward identities
\begin{eqnarray}
k_1^{\mu}\Gamma_{\rho\mu\nu}(q,-k_1,-k_2) &=&
(q^2 g_{\rho\nu} - q_{\rho}q_{\nu}) - 
(k^2_2 g_{\rho\nu} - k_{2\rho}k_{2\nu}) \, ,\nonumber\\
k_1^{\mu}\gamma_{\mu} &=& -i \bigg(S^{-1}(k_1+p)-S^{-1}(p)\bigg)
\, ,
\end{eqnarray}
triggered by the action of $k_1^{\mu}$ (or $k_2^{\nu}$) on 
${{\cal T}_{s}}^{ab}_{\mu\nu}$ and ${{\cal T}_{t}}^{ab}_{\mu\nu}$,
respectively.

After carrying out the above s-cannel -- t-channel cancellation,
one is left with a set of ``pure'' propagator-like contributions 
which define the effective PT   
vacuum polarization of the gluon, 
denoted by $\widehat{\Pi}_{\mu\nu}(q)$,
given by \cite{PT} (Fig.2)
\begin{equation}
\label{PTgg}
\widehat{\Pi}_{\mu\nu}(q)\ =\ \frac{g^2}{16\pi^2}\, \frac{11c_A}{3}\, 
(q^2 g_{\mu\nu}q^2 - q_{\mu}q_{\nu}) \, 
\Big[\, \ln\Big(-\frac{q^2}{\mu^2}\Big)\, +\, C_{UV}\, 
\Big]\, .
\end{equation}
Here, $C_{UV}=1/\epsilon -\gamma_E + \ln 4\pi + C$, where $C$
is a GFP-independent constant and 
$\mu$ is the subtraction point. Notice that, as happens
in QED, $\widehat{\Pi}_{\mu\nu}(q)$ captures
the one-loop leading logarithmic corrections, i.e. the
coefficient $\frac{11c_A}{3}$ multiplying the logarithm coincides
with the coefficient of the one-loop $\beta$ function of quark-less QCD. 

Similarly one may define the GFP-independent one-loop quark-gluon
vertex $\widehat{\Gamma}_{\alpha}^{(1)}(Q,Q')$ (Fig.3).
In addition to being GFP-independent, by virtue of Eq.\ (\ref{FWI})
$\widehat{\Gamma}_{\alpha}^{(1)}(Q,Q')$ satisfies the following QED-like 
Ward-identity
\be
q^{\alpha}\widehat{\Gamma}_{\alpha}^{(1)}(Q,Q')=
\widehat{\Sigma}^{(1)}(Q)-\widehat{\Sigma}^{(1)}(Q'),
\label{WI1}
\ee
where $\widehat{\Sigma}^{(1)}$ is the PT one-loop
quark self-energy, which coincides with the conventional 
one computed in the Feynman gauge.

The construction presented above goes through without major 
conceptual modifications (but with minor operational adjustments) 
in the context of non-Abelian 
gauge theories, such as the electroweak part of the Standard Model,
where the gauge fields have been endowed with masses through the
usual Higgs mechanism.  

\section{Conclusions}

We have seen that  
 the Breit-Wigner resummation formalism can be self-consistently 
extended to the case of non-Abelian gauge theories, provided that
one resorts to the pinch technique rearrangement of the physical
amplitude. To accomplish this one needs invoke only the full
exploitation of the elementary Ward-identities of the theory,
in conjunction with unitarity, analyticity, and renormalization group
invariance. 

From the  phenomenological point of  view the above  framework enables
the  construction of  Born-improved amplitudes  in which  all relevant
physical information has been correctly encoded.  This in turn will be
useful for the detailed study of the physical properties of particles,
most importantly  the correct extraction of their  masses, widths, and
line shapes.

The formalism described in
this paper has been recently extended at the two-loop level
\cite{PT2L}, leading to the exact replication of all the desirable
properties listed above (items [(a)]--[(f)]). 

\newpage

\section*{Acknowledgements}
I would like to thank the organizers, and especially Massimo Porrati,
for creating a very pleasant atmosphere.  
It is a great pleasure to acknowledge numerous stimulating 
discussions with Alberto Sirlin, who has greatly influenced my
current understanding of the various topics presented here.
This work has been funded by CICYT under the Grant AEN-99-0692, by DGESIC under
the Grant PB97-1261 and by the DGEUI of the ``Generalitat Valenciana'' 
under the Grant GV98-01-80.

\vspace{2.5cm}
\centerline{\large {\bf FIGURE CAPTIONS}}
\vspace{0.3cm}

\noindent
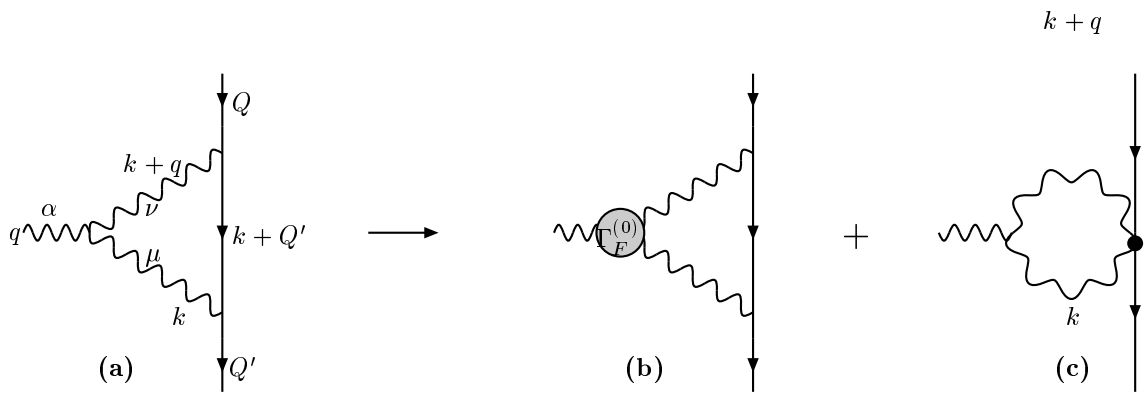
Fig.1: The basic pinch technique rearrangement of the one-loop
vertex (a) into a purely vertex-like piece (b) and a propagator-like
piece (c).

\medskip

\noindent
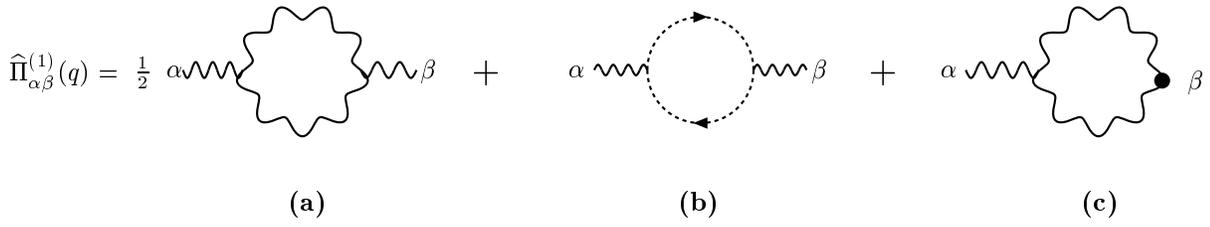
Fig.2: The effective pinch technique gluon self-energy.

\medskip

\noindent
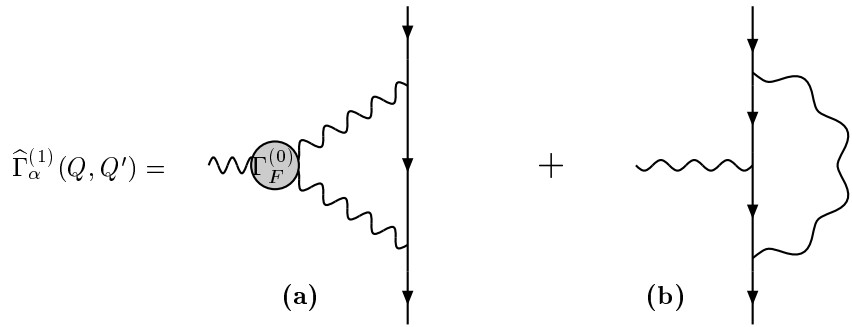
Fig.3: The effective pinch technique one-loop gluon-quark vertex.

\medskip
\noindent
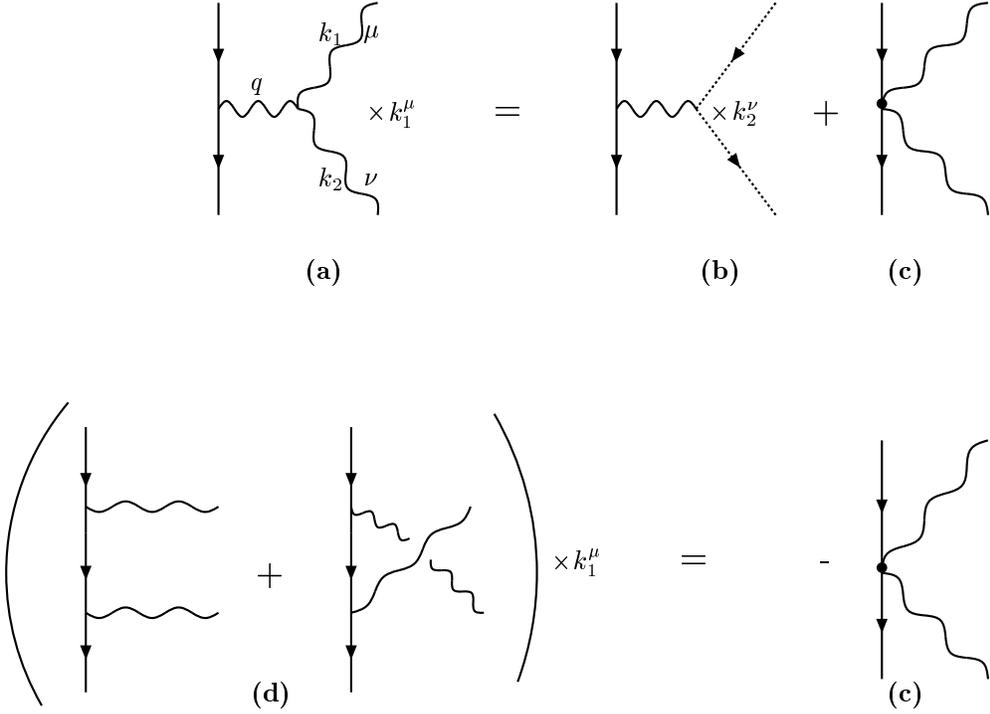
Fig.4: The fundamental BRS-driven $s$-channel -- $t$-channel 
cancellation.

\end{document}